\documentclass{book}

\usepackage{6x8mlt2e}

\input epsf             
\newcommand{\figeps}[1]{\centerline{\epsfbox{#1}\hfill}}
       
\makeindex

\begin{document}
\pagenumbering{roman}
\setcounter{page}{5}

\include {contr}   
\tableofcontents       

\clearpage
\pagenumbering{arabic}
\part{Markov games}                  
      \author{Andrew Allison}
\chapter{State-Space Visualisation and Fractal Properties
   of Parrondo's Games}
\chapterauthors{Andrew Allison\footnote{Centre for Biomedical Engineering (CBME) \& Department of Electrical
and Electronic Engineering, Adelaide University, South Australia,
5005.}\\
Charles Pearce\footnote{Department of Applied Mathematics, Adelaide
  University, South Australia, 5005.}\\
Derek Abbott\footnote{Centre for Biomedical Engineering (CBME) \& Department of Electrical
and Electronic Engineering, Adelaide University, South Australia, 5005.}}
\label{parrondos_fractal}

\section{Introduction}

In Parrondo's games, the apparently paradoxical situation occurs where
individually losing games combine to win
\cite{Harmer1999c,Harmer2000}. 
The basic formulation and definitions of Parrondo's games are
described in Harmer et alii
\cite{Harmer1999a,Harmer1999b,Pearce1999a,Pearce1999b}.
These games have recently gained considerable attention as they are
physically motivated and have been related to physical systems such as
the Brownian ratchet \cite{Harmer1999b}, lattice gas automata
\cite{Meyer2000} and spin systems \cite{Moraal2000}. Various authors have pointed out
interest in these games for areas as diverse as biogenesis \cite{Davies2001}, political models
\cite{Moraal2000}, small-world networks \cite{Toral2001}, economics \cite{Moraal2000}
and population genetics \cite{McClintock1999}.

In this chapter, we will first introduce the relevant properties of
Markov transition operators and then introduce some terminology and
visualisation techniques from the theory of dynamical systems.  We
will then use these tools, later in the chapter, to define and
investigate some interesting properties of Parrondo's games.

We must first discuss and introduce the mathematical machinery, terms
and notation that we will use. The key concepts are~:~
\begin{description}

   \item[state]~:~ This contains all of the information that we need
   to specify what is happening in the system at any given time.

   \item[time-varying probability vector]~:~This is a time-varying
   probability distribution which specifies the probabilities that the
   system will be in certain states and any given time.

   \item[transition matrix]~:~This is a Markov operator which which
   determines the way in which the time varying probability vector
   will evolve over time.

\end{description}
These concepts are defined and discussed at length in many of the standard text books on 
stochastic processes \cite{Karlin1975,Karlin1998,Norris1997,Yates1999}.

Time-homogeneous sequences of regular Markov transition operators have
unique stable limiting state-probabilities.  The state-space
representations of the associated time-varying probability vectors
converge to unique points. If the sequence of Markov transition
operators is not homogeneous in time then the sequence time-varying
probability vectors generated by the products of these different
operators need not converge to a single point, in the original state
space. It is possible to construct quite simple examples to show that
this is the case.

If the sequences are periodic then it is possible to incorporate the
finite memory of these systems into a new definition of ``state.'' The
new inhomogeneous systems can be re-defined as strictly homogeneous
Markov processes.  These new Markov processes, with new states, will
generally have unique limiting probability vectors.

If we allow the sequence to become indefinitely long then the amount
of memory required grows without bound. It is still possible, in
principle, to define these indefinitely long periodic sequences as
homogeneous Markov process although the definition, and encoding, of
the states would require some care.  We can consider any one
indefinite sequence of operators as being one of many possible
indefinite sequences of operators. If we do this then most of the
possible sequences will appear to be ``random.'' We can learn
something about the general case by studying indefinitely long random
sequences.

If the sequence of operators is chosen at random then the time varying
probability vector, as defined in the original state-space, does not
generally converge to a single unique value. Simulations show that the
time-varying probability vector assumes a distribution in the original
state-space which is self-similar, or ``fractal,'' in appearance.  The
existence of fractal geometry is established, with rigor, for some
particular Markov games. We establish a transcendental equation which
allows the calculation of the Hausdorff dimensions of these fractal
objects.

If state-transitions of the time-inhomogeneous Markov chains are
associated with rewards then it is possible to show that even simple,
``two-state,'' Markov chains can generate a Parrondo effect, as long
as we are free to choose the reward matrix.  Homogeneous sequences of
the individual games generate a net loss over time. Inhomogeneous
mixtures of two games can generate a net gain.

We show that the expected rates of return, or moments of the reward
process, for the time-inhomogeneous games are identical to the
expected rates of return from a homogeneous sequence of a
time-averaged game. This is a logical consequence of the Law of Total
Probability and the definition of expected value.

Two different views of the time-inhomogeneous process emerge,
depending of the viewpoint that one takes:
\begin{itemize}

\item If you have access to the
history of the time-varying probability vector and you have a memory to
store this information and you choose to represent this data in
state-space then you will see distributions with fractal
geometry. This is more or less the view that a large casino might have
if they were to visualise the average states of their many customers.

\item If you do not have access to the time-varying probability vector or 
you have no memory in which to store this information then all that
you can see is a sequence of rewards from a stochastic process. The
internal details of this process are hidden from you. You have no way
of knowing precisely how this process was constructed from an
inhomogeneous sequence of Markov operators. There is no experiment
that you can perform to distinguish between the time-inhomogeneous
process and the time-averaged process. The time-averaged process is
a homogeneous sequence of a single operator. We can calculate a
single unique limiting value for the probability vector. This is more
or less the view that a single, mathematically inclined, casino patron
might have if they were playing against some elaborate poker
machine. The internal workings of the machine would be hidden from the
customer but it would be possible to perform some analysis of the
outcomes and form an estimate of the parameters for the time-averaged
model.

\end{itemize}

We show that the time inhomogeneous process is consistent in the sense
that the ``casino'' and the ``customer'' will always agree on the
expected winnings or losses of the customer. In more technical terms,
the time-average, which the customer sees, is the same as the
ensemble-average over state-space, which the casino can calculate.

\section{Time-Homogeneous Markov Chains and Notation}
Finite discrete-time Markov chains can be represented in terms of
matrices of conditional transition probabilities.  These matrices are
called Markov transition operators. We denote these by capital letters
in brackets, eg~:~$\left[ A \right]$ where $A_{i,j} = \mbox{Pr}
\left\{K_{t+1} = j | K_t = i \right\}$ and $K \in \mathcal{Z}$ is some
measure of displacement or the ``state'' of the system.  The Markov
property requires that $A_{i,j} $ cannot be a function of $K$ but it
can be a function of time, $t$. In Parrondo's original games, $K$,
represents the amount of capital that a player has.  There is a one-to-one 
mapping between Markov games and the Markov transition operators
for these games.  We will refer to the games and the transition
operators interchangeably.

The probability that the system will be in any one state at a given
instant of time can be represented by a distribution called the
time-varying probability vector.  We represent this probability mass
function, at time $t$, using a row vector, ${\bf V_t}$. We can represent the 
evolution of the Markov chain in time using a simple Matrix equation,
\begin{equation}
  {\bf V_{t+1}}  = {\bf V_t} \cdot \left[A \right]~.
  \label{eq:time_evolution}
\end{equation}
This can be viewed as a multi-dimensional finite difference
equation. The initial value problem can be solved using generating
function, or Z transform, methods.
Sequences of identical Markov transition operators,
where $\left[ A \right]$ does not vary, are said to be
time-homogeneous.  A Markov transition operator is said to be regular
if some positive power of that operator has all positive elements.
Time-homogeneous sequences of regular Markov transition operators
always have stable limiting probability vectors, 
$\lim_{t\rightarrow \infty } \left( {\bf V_t} \right)
={\bf \Pi}$.
The time varying probability vector reliably converges to a single point
\cite{Karlin1975,Karlin1998,Norris1997,Yates1999}.

We can think of the space which contains the time-varying probability
vectors, and the stable limiting probability vector, as a vector space
which has a strong analogy to the state-space which is used in the
theory of control.  We shall refer to this space as ``state-space,''
$\left[ 0 , 1 \right] ^ N$,
and we will refer to the time-varying probability vector
as a state vector.  This terminology is used in the engineering 
literature \cite{Yates1999}.  We emphasise that the 
``{\it state-vector'},''  ${\bf V_t} \in {\Re}^N$ is distinct
from the ``{\it state'}'' of the system, $K \in \mathcal{Z}$, used in Markov chain terminology.
As a simple example, we can 
consider the regular Markov transition operator
\begin{equation}
\left[A \right] =
\left[
\begin{array}{cc}
\frac{13}{16} & \frac{3}{16} \\
\frac{1}{16} & \frac{15}{16} \\
\end{array}
\right]
\label{eq:transition_A1}
\end{equation}
using the initial condition
\begin{equation}
  {\bf V_t} = \left[ V_0 , V_1 \right]~
             = \left[  \frac{3}{4} ,  \frac{1}{4}  \right] \mbox{~when~} t = 0 .
  \label{eq:initial_condition_A1}
\end{equation}

The components of ${\bf V_t}$ are $V_0$ and $V_1$ and these can
be considered to be the dimensions of the Cartesian space which we
call ``state-space''. This space has a clear analogy with the
phase-space of Poincare and the state-space used in the theory of
control.  It also has some analogy with the ``$\gamma$'' or gaseous
phase-space of Gibbs and the phase-space used in Lagrangian dynamics
although we must be careful not to press these analogies too far since
the state-spaces of physics and of Markov chains use different
transition operators which obey different conservation laws.

\begin{figure}[h!]
\hskip.5in   \figeps{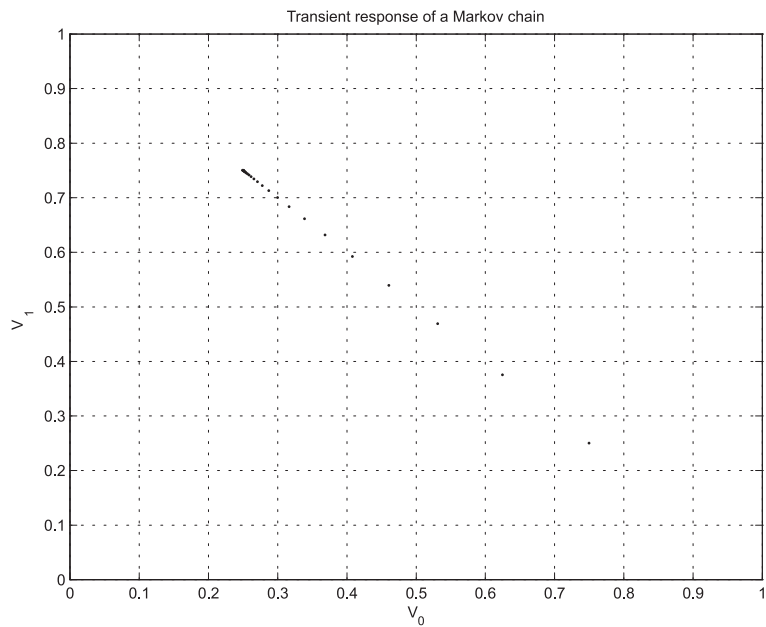}
   \caption{State-space trajectory of a Markov chain
\label{fig:transient}}
\end{figure}

A fundamental question in the study of dynamical systems is to
classify how they behave as $t \rightarrow \infty$ and all transient
effects have decayed.
The evolution of the state vector of a
discrete-time Markov chain generally traces out a sequence of points
or ``trajectory'' in the state-space.
The natural technique would be to draw a graph of this trajectory.  
As an example of this, we can consider the trajectory of 
the time homogeneous Markov chain, described by Equations
\ref{eq:transition_A1} and \ref{eq:initial_condition_A1}, which is shown in Figure
\ref{fig:transient}.

The state vector, ${\bf V_t}$, always satisfies the constraint,
$V_0 + V_1 = 1$. This follows from the law of total probability.  The
state-vector is always constrained to lie within an $N-1$ dimensional
subspace of the $N$ dimensional state-space. The dynamics of the
system all occur within this sub-space. This is clearly visible in
Figure \ref{fig:transient}. We can think of the set
\begin{equation}  
   M = \left\{ \left[ V_0, V_1 \right] ~|~ 
   (0 \leq V_0 \leq 1 ) \wedge  (0 \leq V_1 \leq 1 ) \wedge  (V_0 + V_1 = 1) \right\}~,
   \label{eq:state-manifold}
\end{equation}
as a state manifold for the dynamical system defined by Equations
\ref{eq:transition_A1} and \ref{eq:initial_condition_A1}.  The state
manifold has a dimension which is smaller than the embedding state-space.
This is a result of the fact that there is a conservation law (the law
of total probability) which constrains the dynamics of the system.
For this example, the
sequence converges to a stable fixed point at 
${\bf \Pi}= \left[ \frac{1}{4} , \frac{3}{4} \right]$.
It can be shown that sequences of this type always converge to single
stable fixed points as long as the Markov transition operators are
regular and time-homogeneous
\cite{Karlin1975,Karlin1998,Norris1997,Yates1999}.
The convergent points are the appropriate state-space representation
of the stable limiting probabilities for the Markov chain.

\section{Time-Inhomogeneous Markov Chains}
The existence, uniqueness and dynamical stability of the fixed point
are important parts of the theory of Markov chains but we must be
careful not to apply these theorems to systems where the basic
premises are not satisfied.  If the Markov transition operators are
not homogeneous in time then there may no longer a single fixed point
in state-space. The state vector can perpetually move through two or
more points without ever converging to any single stable value.  To
demonstrate this important point, we present a simple example, using
two regular Markov transition operators~:
\begin{equation}
\left[S \right] =
\left[
\begin{array}{cc}
\frac{3}{4} & \frac{1}{4} \\
\frac{3}{4} & \frac{1}{4} \\
\end{array}
\right]
\label{S1}
\end{equation}
and
\begin{equation}
\left[T \right] =
\left[
\begin{array}{cc}
\frac{1}{4} & \frac{3}{4} \\
\frac{1}{4} & \frac{3}{4} \\
\end{array}
\right]~.
\label{T1}
\end{equation}

The rows of these matrices are all identical. This indicates that the
outcome of each game is completely independent of the initial state.
The limiting stable probabilities for these regular Markov transition
operators are
${\bf \Pi_S} = \left[ \frac{3}{4} , \frac{1}{4} \right] $ and 
${\bf \Pi_T} = \left[ \frac{1}{4} , \frac{3}{4} \right] $
respectively.  The time-varying probability vector immediately moves
to the stable limiting value after even a single play of each game.
\begin{equation}
   \left[ Q \right] \cdot \left[ S \right] = \left[ S \right]
   \label{eq:single_step_1}
\end{equation}
and
\begin{equation}
   \left[ Q \right] \cdot \left[ T \right] = \left[ T \right]
   \label{eq:single_step_2}
\end{equation}
for any conformable stochastic matrix $\left[ Q \right]$.
This leads to some interesting corollaries:
\begin{equation}
   \left[ T \right] \cdot \left[ S \right] = \left[ S \right]
   \label{eq:single_step_TS}
\end{equation}
and
\begin{equation}
   \left[ S \right] \cdot \left[ T \right] = \left[ T \right]
   \label{eq:single_step_ST}
\end{equation}
If we play an indefinite alternating sequence of these games,
$\{ S T S T \cdots \}$,
then there are two simple ways in which we can associatively group the
terms:
\begin{eqnarray}
   {\bf V_{2N}} & = & {\bf V_{0}} 
    \left( \left[ S \right]  \left[ T \right] \right)
    \left( \left[ S \right]  \left[ T \right] \right)
    \cdots
    \left( \left[ S \right]  \left[ T \right] \right) \\
    & = &   {\bf V_{0}}  \left[ T \right] \\
    & \Rightarrow & {\bf \Pi} = {\bf \Pi_{T}} 
   \label{eq:even_grouping}
\end{eqnarray}
and
\begin{eqnarray}
   {\bf V_{2N+1}} & = & 
    \left( {\bf V_{0}}   \left[ S \right] \right)
    \left( \left[ T \right]  \left[ S \right] \right)
    \left( \left[ T \right]  \left[ S \right] \right)
    \cdots
    \left( \left[ T \right]  \left[ S \right] \right) \\
    & = &   {\bf V_{0}}  \left[ S \right] \\
    & \Rightarrow & {\bf \Pi} = {\bf \Pi_{S}}~. 
   \label{eq:odd_grouping}
\end{eqnarray}

If we {\it assume} that there is a unique probability limit then we
must conclude that $ {\bf \Pi_{S}} = {\bf \Pi_{T}}$ and hence
$\frac{1}{4} = \frac{3}{4}$ which is a contradiction. We can invoke
the principle of excluded middle (reductio ad absurdum) to conclude
that the assumption of a single limiting stable value for 
$ \lim_{t \rightarrow \infty} \left( {\bf V_t} \right) $ is false.  In the
asymptotic limit as $t \rightarrow \infty$, the state vector
alternately assumes one of the {\bf {\it two}} values ${\bf \Pi_S}$ or
${\bf \Pi_T}$. We refer to the set of all recurring state vectors of
this type, $\{ {\bf \Pi_S} , {\bf \Pi_T} \}$, as the {\it attractor}
of the system. In more general terms an attractor is a set of points
in the state-space which is invariant under the system dynamics in the
asymptotic limit as 
$t \rightarrow \infty$.

\subsection{Reduction of the periodic case to a Time-Homogeneous Markov Chain}

In the last section, we considered a short sequence of length 2. This
can be generalised to an arbitrary length, $N \in \mathcal{Z}$. It is
possible to associatively group the operators into sub-sequences of
length $N$. As with the sequences of length two, the choice of time
origin is not unique. We are free to make an arbitrary choice of time
origin with the initial condition at $t = 0$. We can think of the
operators as having an offset of $ n \in \mathcal{Z}$, where $0 \leq n \leq N-1 $
within the sub-sequence. We can also calculate a new equivalent
operator to represent the entire sequence, $A_{\mbox{eq}} = \prod_{n
= 0}^{N-1} A_n$. We can then calculate the steady-state probabilities
associated with this operator, 
${\bf  \Pi_{ \mbox{eq} } } = {\bf  \Pi_{ \mbox{eq} } } \cdot A_{\mbox{eq}} $. 
We can refer the asymptotic trajectory of the
time varying probability vector to this fixed point,
${\bf  V_{ ( t \pmod{N} ) } }  = {\bf  \Pi_{\mbox{eq} }  } \cdot  \Pi_{n = 0}^{  (t \pmod{N} ) -1 } A_n $.
In the periodic case, there is generally not a single fixed point in
the original state-space but the time varying probability vector
settles into a stable limit cycle of length $N$. If we aggregate time,
modulo $N$ then we can re-define what we mean by ``state'' and we can
define a new state-space in which the time-varying vector does
converge to a single point.

If we allow the length of the period, $N$, to become indefinitely long
$N \rightarrow \infty$ then our new definition of ``state'' becomes
infinitely complicated. We would have to contemplate indefinitely
large offsets, $n \rightarrow \infty$, within the infinitely long
cycle.  If we wish to avoid the many paradoxes that infinity can
conceal then we really should consider the case with ``infinite''
period as being qualitatively different from the case with finite
period, $N$.

\subsection{Random Selection of Markov transition operators}

\section{Two simple Markov Games that Generate a Simple Fractal in
  State-Space}

\begin{figure}[h!]
\hskip.5in   \figeps{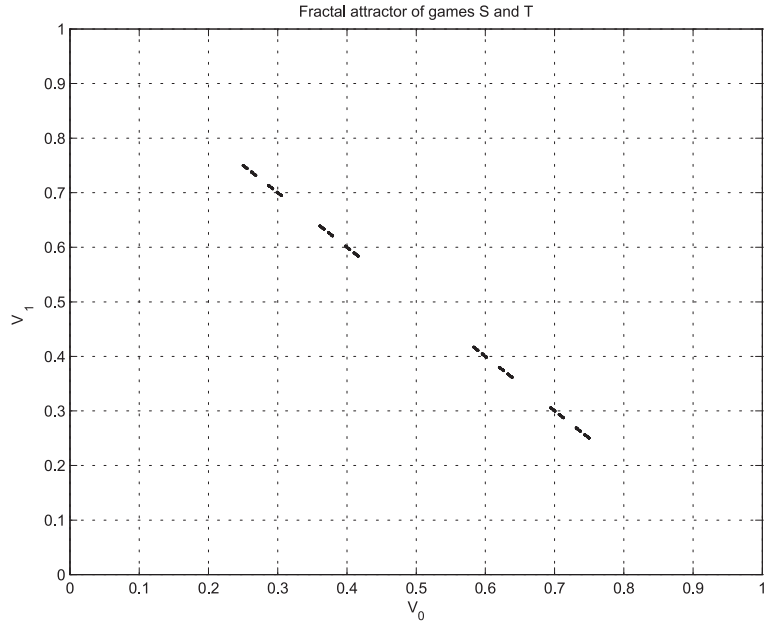}
   \caption{A fractal attractor generated by games S and T
\label{fig:cantor_fractal} }
\end{figure}

We proceed to construct a simple system in which operators are
selected at random and we will use the standard theories regarding
probability and expected values to derive some useful results.
If we modify the system specified by Equations
\ref{S1} and \ref{T1}~:
\begin{equation}
\left[S \right] =
\left[
\begin{array}{cc}
\frac{5}{6} & \frac{1}{6} \\
\frac{1}{2} & \frac{1}{2} \\
\end{array}
\right]
\label{S2}
\end{equation}
and
\begin{equation}
\left[T \right] =
\left[
\begin{array}{cc}
\frac{1}{2} & \frac{1}{2} \\
\frac{1}{6} & \frac{5}{6} \\
\end{array}
\right]
\label{T2}
\end{equation}
and select the sequence of transition operators at random 
then the attractor becomes an infinite set.
If we were to play a homogeneous sequence of either of these games 
then they would have the same stable limiting probabilities
as before, ${\bf \Pi_S}$ and ${\bf \Pi_T}$, and the dynamics would be 
similar to those shown in Figure \ref{fig:transient}.
In contrast, if we play an indefinite {\bf {\it random}} sequence 
of the new games S and T,
$\{ S T S S T S T T S T T \cdots \}$, then 
there are no longer any stable limiting probabilities 
and the attractor has a fractal or ``self-similar'' appearance 
which is shown in Figure \ref{fig:cantor_fractal}.

\subsection{The Cantor Middle-Third Fractal}
\label{sec:Middle-Third}

These games have been constructed in such a way that they
generate the Cantor middle-third fractal. 

It should be noted that the Cantor Middle-Third fractal is an
uncountable set and so a, countably infinite, random sequence of
operators will ever generate enough points to cover the entire
set. The solution to this problem is to consider the uncountably
infinite set generated by all possible infinite, random sequences of
operators. We can construct a probability measure on the resulting set
and then we can calculate probabilities and expected values. It is also
reasonable to talk about the probability density function of the
time-varying probability vector in the state-space.

\begin{figure}[h!]
\hskip.5in   \figeps{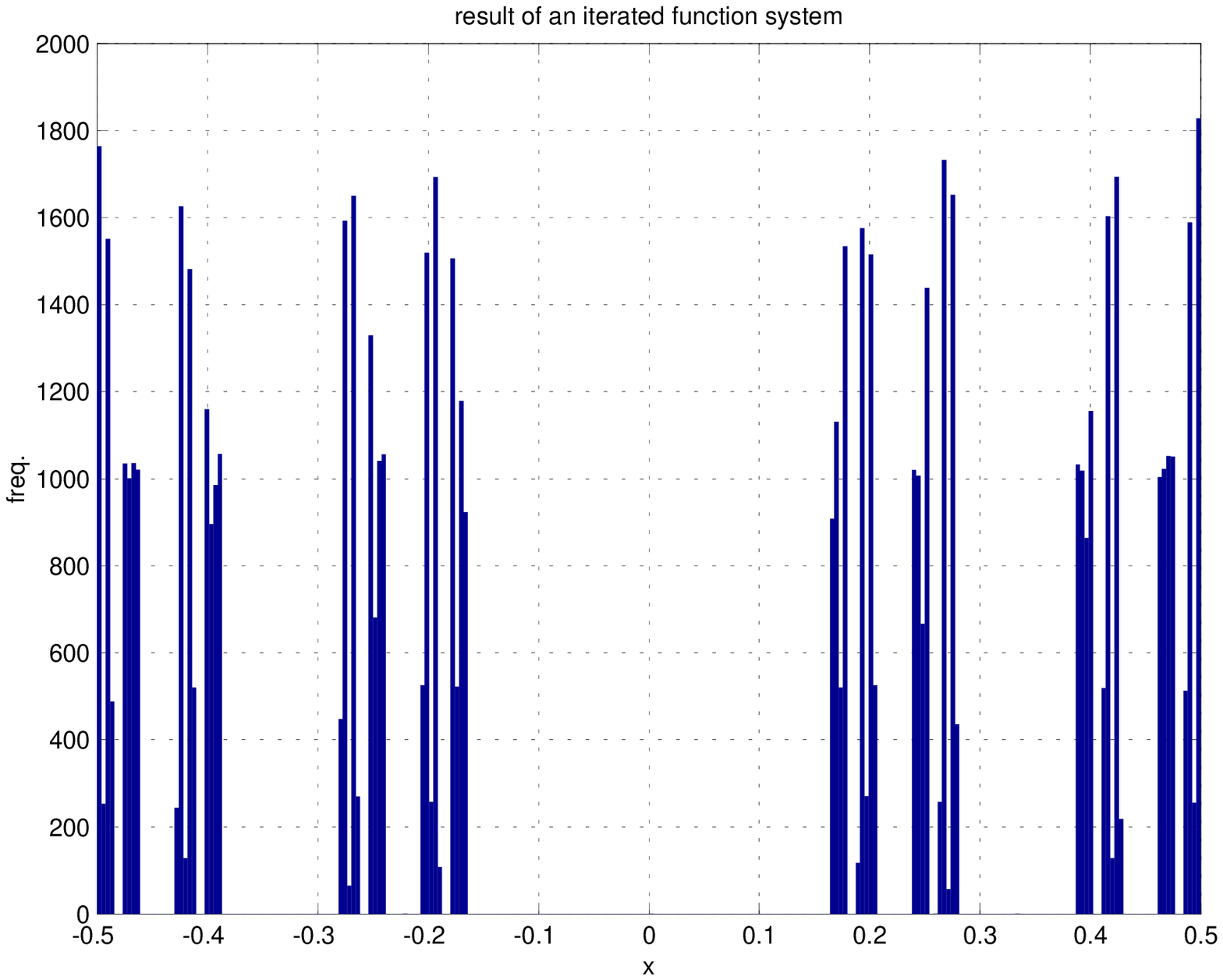}
   \caption{A histogram of the distribution of ${\bf V_t}$ in
     state-space
\label{fig:cantor_histogram }}
\end{figure}

In order to stimulate intuition, we can simulate the process and
generate a histogram, showing the distribution of the time varying
probability vector. The result is shown in Figure \ref{fig:cantor_histogram }.
For the $x$ axis in this figure, we {\bf{\it could}} have chosen the
first element of the time varying probability vector, $V_0$ but this would
not have been the easiest way to analyse the dynamics.
It is better if we choose another parameterization. If we
examine the eigenvectors of the matrices in Equations \ref{S2} and
\ref{T2} then we find that a better re-parameterization is:
\begin{equation}
   x = V_0 - V_1
   \label{eq:define_x}
\end{equation}
and
\begin{equation}
   y = V_0 + V_1~.
   \label{eq:define_y}
\end{equation}
Of course, we always have $y = 1$ and $x$ is a new variable in the range
$ -1 \leq x \leq +1 $. The Cantor fractal lies in the unit interval
$ -\frac{1}{2}  \leq x \leq \frac{1}{2}$ which is the $x$ interval shown in 
Figure \ref{fig:cantor_histogram }.
The transformation for matrix $\left[ S \right]$, in Equation
\ref{S2} reduces to:
\begin{equation}
   \left( + \frac{1}{2} - x_{t+1} \right) = \frac{1}{3} \cdot \left( + \frac{1}{2} - x_{t} \right)
   \label{eq:new_S}
\end{equation}
and the transformation for matrix $\left[ T \right]$, in Equation \ref{T2} reduces to:
\begin{equation}
   \left( - \frac{1}{2} - x_{t+1} \right) = \frac{1}{3} \cdot \left( - \frac{1}{2} - x_{t} \right)~.
   \label{eq:new_T}
\end{equation}
The transformation $S$ has a fixed point at $x = + \frac{1}{2}  $ and the transformation $T$
has a fixed point at $x = - \frac{1}{2} $. If we choose these transformations as random then the recurrent values of $x$ lie in the interval between the fixed points, $ -\frac{1}{2}  \leq x \leq \frac{1}{2}$.
This is precisely the iterated function system for the Cantor Middle-Third Fractal.
These are described in Barnsley \cite{Barnsley1988}.  

The most elementary analysis that we can perform is to calculate the
dimension of this set. If we assume conservation of measure then every
time we perform a transformation, we reduce the diameter by a factor
of $\frac{1}{3}$ but the transformed object is geometrically half of
the original object so we can write
\begin{equation}
   \frac{1}{2} = (\frac{1}{3})^D
   \label{conservation-of_measure}
\end{equation}
where $D$ is the fractional dimension. 
This is the law of conservation of measure for this particular system.
We can solve this equation for $D$ to get $D = \frac{\log(2)}{\log(3)} \approx 0.630929 \cdots $.

We can invert the rules described in Equations \ref{eq:new_S} and \ref{eq:new_T} giving:
\begin{equation}
   x_{t} = 3  x_{t+1} - 1
   \label{eq:new_S_inv}
\end{equation}
and
\begin{equation}
   x_{t} = 3  x_{t+1} + 1~.
   \label{eq:new_T_inv}
\end{equation}
If we consider these equations, together with the law of conservation
of total probability then we get a self-similarity rule for the PDF
(or Probability Density Function), $p(x)$, of the time varying probability vector,
${\bf V_{t}}$~:
\begin{equation}
   \frac{3}{2} p \left( 3 x - 1 \right) +
   \frac{3}{2} p \left( 3 x + 1 \right) =
   p \left( x \right)~.
   \label{eq:self_similar_PDF}
\end{equation}
This PDF, $p(x)$ is the density function towards which the histogram in Figure
\ref{fig:cantor_histogram } would converge if we could collect enough samples.
The self-similarity rule for the PDF gives rise to a recursion rule for the
moment generating function,~$\Phi \left( \Omega \right) = E \left( e^{j \Omega x} \right)$~:
\begin{equation}
   \Phi \left( \Omega \right) =
    \Phi \left( \frac{\Omega}{3} \right) \cdot \cos \left( \frac{\Omega}{3}  \right)~.
   \label{eq:self_similar_GF}
\end{equation}
We can evaluate the derivatives at $\Omega = 0$ and calculate as many
of the moments as we wish. We can calculate the mean, $\mu$, and
the variance $\sigma^2 $~: 
\begin{eqnarray}
   \mu     & = & 0 \\
   \label{fractal_mean}
   \sigma^2  & = & \frac{1}{8}
   \label{eq:fractal_variance}
\end{eqnarray}
These algebraic results are consistent with results from numerical
simulations.

\subsection{Iterated Function Systems}

The cause of the fractal geometry is best understood if we realise
that Markov transition operators perform affine transformations on the
state-space.  An indefinite sequence of different Markov transition
operators is equivalent to an indefinite sequence of different affine
transformations which is called an ``Iterated Function System''.  We
refer the reader to the work of Michael Barnsley \cite{Barnsley1988}
and the theory of Iterated Function Systems to show that fractal
geometry is quite a general property of a system of randomly selected
affine transformations.

\section{An Equivalent Representation of the Random Selection of Markov Transition Operators}
\begin{figure}[h!]
\hskip.5in   \figeps{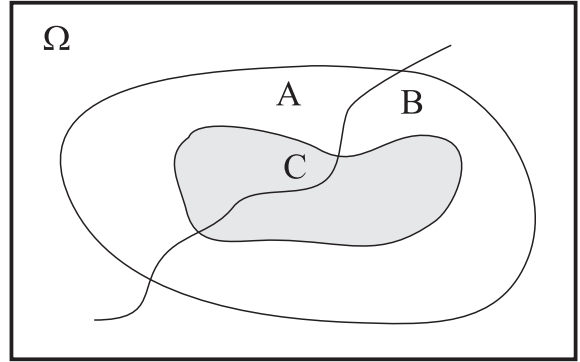}
   \caption{Set Relationships and Change of Probability
\label{fig:set_relations}}
\end{figure}
Consider two mutually exclusive events, $A \cap B = \emptyset$,
embedded within some probability space $\left( \Omega, \mathcal{F}, P
\right)$.  Consider any third event $C \subseteq A \cup B$.  These
events are represented in Figure \ref{fig:set_relations}.
The law of total probability asserts that
\begin{equation}
        \mbox{Pr}(C) =  \mbox{Pr}(C|A) \cdot \mbox{Pr}(A) +  \mbox{Pr}(C|B) \cdot  \mbox{Pr}(B)~.
   \label{eq:total_probability}
\end{equation}
We can now make the following particular identifications:
\begin{eqnarray}
   C & \equiv & \left\{ X \in \Omega~|~K_{t+1} = i ~ \wedge ~ K_{t} = j\right\} \\
   A & \equiv & \left\{ \mbox{played game} ~ A \right\} \\        
   B & \equiv & \left\{ \mbox{played game} ~ B \right\}~.          
\end{eqnarray}
If we select games $A$ and $B$ at random with probabilities of
$\gamma$ and $(1-\gamma)$ respectively then we can write 
$\mbox{Pr}(A) = \gamma$ and $\mbox{Pr}(B) = (1-\gamma)$~.
By definition, the Markov matrices for games $A$ and $B$ contain conditional probabilities for state transitions~:
\begin{eqnarray}
A_{i,j} & = & \mbox{Pr}\left\{ \left( K_{t+1} = j ~|~ K_t = i \right)   
            ~\wedge~  \mbox{played game} ~ A \right\} \\
B_{i,j} & = & \mbox{Pr}\left\{ \left( K_{t+1} = j ~|~ K_t = i \right)   
            ~\wedge~  \mbox{played game} ~ B \right\}~.
\end{eqnarray}
Note that in this case $C = A \cup B$.
We can define a new operator corresponding to the events $C_{i,j}$~:
\begin{equation}
   C_{i,j} = \mbox{Pr}\left\{  K_{t+1} = j ~|~ K_t = i  \right\} 
\end{equation}
and Equation \ref{eq:total_probability} reduces to
\begin{equation}
  C_{i,j} = A_{i,j} \cdot \gamma +  B_{i,j} \cdot \left( 1 - \gamma \right)~.
  \label{eq:time_average_game}
\end{equation}
The conditional probabilities of state transitions of the
inhomogeneous Markov process generated by games $A$ and $B$ are the
same as the conditional probabilities of a new equivalent game called
``Game $C$.'' The transition matrix for Game $C$ is a linear convex
combination of the matrices for the original basis games, $A$ and $B$.
Even if we have complete access to the state of the system then there
is no function that we can perform on the state, or state transitions,
which could allow us to distinguish between a homogeneous sequence of
Games $C$ and an inhomogeneous {\bf{\it random}} sequence of Games $A$
and $B$. We refer to game $C$ as the time-average model.  This is
analogous to the state-space averaged model found in the theory of
control \cite{middlebrook1976}.

\section{The Phenomenon of Parrondo's Games}

\subsection{Markov Chains with Rewards}

\begin{figure}[h!]
\hskip.5in   \figeps{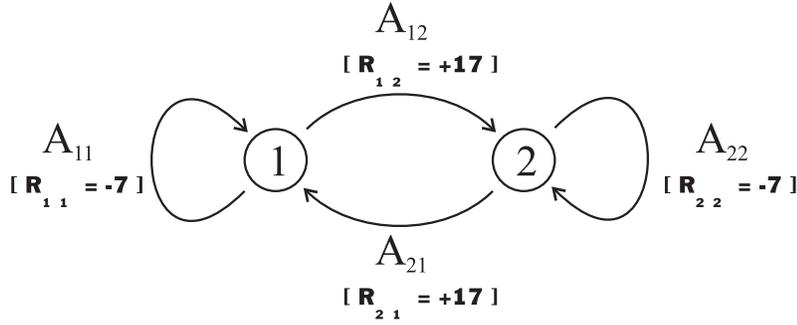}
   \caption{State Transition Diagram for ``Game A'' with rewards.
\label{fig:two_state_transition}}
\end{figure}

Suppose that we apply a reward matrix to the process:
\begin{equation}
   R_{i,j} = \mbox{reward if}~  ( K_{t+1} = j ) ~|~ ( K_t = i )~.
\end{equation}
There is a specific reward associated with each specific state
transition.  We can think of $R_{i,j}$ as the reward that we earn when
a transition occurs from state $i$ to state $j$. The state transitions,
rewards and probabilities of transition, for ``Game A'' are shown in
Figure \ref{fig:two_state_transition}.
The state transition diagrams for ``Game B'' and the time averaged
``Game C'' would have identical topology and have identical reward
structure, although the probabilities of transition between states
would be different.  Systems of this type have been analysed by Howard
\cite{Howard1960} although we use different, matrix, notation to perform the
necessary multiplications and summations.

The expected reward from each transition of the time-averaged homogeneous process is :
\begin{equation}
   Y_{i,j} = E\left[  R_{i,j} \cdot C_{i,j} \right]~.
   \label{eq:expected_return_1}
\end{equation}
If we wish to calculate the mean expected reward then we must sum over
all recurrent states in proportion to their probability of occurrence.
This will be a function of the transition matrix, $C$, and the
relevant steady state probability vector, ${\bf \Pi_{C}}$ ~:
\begin{equation}
   Y \left(C \right) = {\bf \Pi_{C}} \cdot  \left(  \left[  R \right]  \circ \left[  C \right] \right)\cdot
                {\bf U}^{T}
   \label{eq:expected_return_2}
\end{equation}
where ``$\circ$'' represents the Hadamard, or element by element,
product and $ {\bf U}^{T}$ is a unit column vector of dimension
$N$. Post-multiplication by ${\bf U}^{T}$ has the effect of performing
the necessary summation. We recall that ${\bf \Pi_{C}}$ represents the
steady state probability vector for matrix $C$.  The function $Y
\left(C \right)$ represents the expected asymptotic return, in units
of ``reward,'' per unit time when the the games are played.

If we include the definition of $C$ in Equation
\ref{eq:time_average_game} in Equation \ref{eq:expected_return_1}
then we can write~:
\begin{eqnarray}
  Y_{i,j} & = &  E\left[  R_{i,j} \cdot \left( \gamma        A_{i,j} + 
                                  \left( 1- \gamma \right) B_{i,j}  \right) \right] \\
          & = &  \gamma             E\left[  R_{i,j} \cdot A_{i,j} \right] + 
                 \left( 1- \gamma \right) E\left[  R_{i,j} \cdot B_{i,j} \right]~.
   \label{eq:linear_decomposition}   
\end{eqnarray}
We can also define~:
\begin{equation}
  Y(A) = {\bf \Pi_{A}} \left( \left[R \right] \circ \left[ A \right] \right)  {\bf U}^{T}
   \label{eq:YA}
\end{equation}
and
\begin{equation}
  Y(B) = {\bf \Pi_{B}} \left( \left[R \right] \circ \left[ B \right] \right)  {\bf U}^{T}
   \label{eq:YB}
\end{equation}
and we might {\bf {\it falsely}} conclude that
\begin{equation}
   Y(C) = \gamma Y(A) +  \left( 1- \gamma \right) Y(B)~.
   \label{eq:FALSE1}
\end{equation}
This would be equivalent to saying that~:
\begin{equation}
  Y(C) = \gamma \left(  {\bf \Pi_{A}} \left( \left[R \right] \circ \left[ A \right] \right)  {\bf U}^{T} \right) + 
         \left( 1 - \gamma \right) \left( {\bf \Pi_{B}}   \left( \left[R \right] \circ \left[ B \right] \right)   {\bf U}^{T} \right)~.
   \label{eq:FALSE2}
\end{equation}
but these equations \ref{eq:FALSE1} and \ref{eq:FALSE2} are in {\bf
error} because Equation
\ref{eq:linear_decomposition} must be summed over all of the
recurrent states of the {\bf {\it mixed}} inhomogeneous games but in
the {\bf false} Equation \ref{eq:FALSE2}, the first term is summed with
respect to the recurrent states of Game ``A'' and the second term is
summed with respect to the recurrent states of game ``B.''
This is an error.
The dependency on state makes the reward process non-linear.
The correct expression for $ Y(C)$ would be~:
\begin{equation}
  Y(C) = \gamma \left(  {\bf \Pi_{C}} \left( \left[R \right] \circ \left[ A \right] \right)  {\bf U}^{T} \right) + \left( 1 - \gamma \right) \left( {\bf \Pi_{C}}   \left( \left[R \right] \circ \left[ B \right] \right)   {\bf U}^{T} \right)~.
\label{eq:TRUE1}
\end{equation}
The difference between the intuitively appealing but {\bf false} Equations 
\ref{eq:FALSE1} and \ref{eq:FALSE2}
and the correct Equation \ref{eq:TRUE1} is the cause of ``Parrondo's paradox.''

\subsection{Parrondo's Paradox Defined}

The essence of the problem is that when we say that ``Game A is
losing'' or ``Game B is losing'' we perform summation with respect to
the steady state probability vectors for Games ``A'' and ``B''
respectively. When we say that ``a random sequence of games A and B is
winning,'' we perform the summation with respect to the steady state
probability vector for the time-averaged game, Game ``C.''

We can say that the ``paradox'' exists whenever we can find two games
$A$ and $B$ and a reward matrix $R$ such that~:
\begin{equation}
   Y\left( \gamma A + \left( 1- \gamma \right) B  \right) \neq \gamma Y(A) +  \left( 1- \gamma \right) Y(B)~.
   \label{eq:Parrondo_definition}
\end{equation}
The ``paradox'' is equivalent to saying that the reward process is not
a linear function of the Markov transition operators.

\subsection{A simple ``Two-State'' Example of Parrondo's Games}
\label{sec:Two-State}
\begin{figure}[h!]
\hskip.5in   \figeps{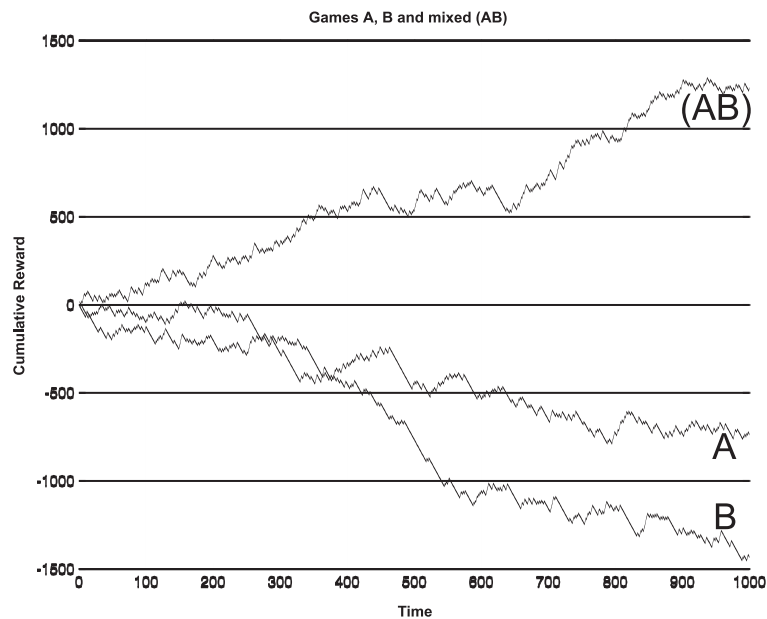}
   \caption{Simulation of a Two-State version of Parrondo's games 
\label{fig:two_state_parrondo}}
\end{figure}

We can show that Parrondo's paradox does exist by constructing a simple example.
We can define 
\begin{equation}
\left[A \right] =
\left[
\begin{array}{cc}
\frac{5}{6} & \frac{1}{6} \\
\frac{1}{2} & \frac{1}{2} 
\end{array}
\right]
\label{eq:A2}
\end{equation}
and
\begin{equation}
\left[B \right] =
\left[
\begin{array}{cc}
\frac{1}{2} & \frac{1}{2} \\
\frac{1}{6} & \frac{5}{6}
\end{array}
\right]
\label{eq:B2}
\end{equation}
The steady state probability vectors are: ${\bf \Pi_A} = \left[
\frac{3}{4} , \frac{1}{4} \right] $ and ${\bf \Pi_B} = \left[
\frac{1}{4} , \frac{3}{4} \right] $.  These games are the same as
games ``S'' and ``T'' defined earlier but we analyse them using the
theory of Markov chains with rewards.
We can define a reward matrix
\begin{equation}
\left[R \right] =
\left[
\begin{array}{cc}
  -7 & +17  \\
 +17 &  -7
\end{array}
\right]
\label{eq:reward}
\end{equation}
and we can apply Equations, 
\ref{eq:YA},  \ref{eq:YB} and \ref{eq:TRUE1}
to get~:
\begin{equation}
Y(A) = \left[
\begin{array}{cc}
\frac{3}{4} & \frac{1}{4}
\end{array} 
\right]
\left(
\left[
\begin{array}{cc}
  -7 & +17  \\
 +17 &  -7
\end{array}
\right]
\circ
\left[
\begin{array}{cc}
\frac{5}{6} & \frac{1}{6} \\
\frac{1}{2} & \frac{1}{2} 
\end{array}
\right]
\right)
\left[
\begin{array}{c}
1 \\
1  
\end{array}
\right]
=~ -1
\label{eq:reward_A}
\end{equation}
and
\begin{equation}
Y(B) = \left[
\begin{array}{cc}
\frac{1}{4} & \frac{3}{4}
\end{array} 
\right]
\left(
\left[
\begin{array}{cc}
  -7 & +17  \\
 +17 &  -7
\end{array}
\right]
\circ
\left[
\begin{array}{cc}
\frac{1}{2} & \frac{1}{2} \\
\frac{1}{6} & \frac{5}{6} 
\end{array}
\right]
\right)
\left[
\begin{array}{c}
1 \\
1  
\end{array}
\right]
=~-1
\label{eq:reward_B}
\end{equation}
and, for the time-average we get~:
\begin{equation}
Y\left( C \right) = \left[
\begin{array}{cc}
\frac{1}{2} & \frac{1}{2}
\end{array} 
\right]
\left(
\left[
\begin{array}{cc}
  -7 & +17  \\
 +17 &  -7
\end{array}
\right]
\circ
\left[
\begin{array}{cc}
\frac{2}{3} & \frac{1}{3} \\
\frac{1}{3} & \frac{2}{3} 
\end{array}
\right]
\right)
\left[
\begin{array}{c}
1 \\
1  
\end{array}
\right]
=~+1~.
\label{eq:reward_C}
\end{equation}
Games ``A'' and ``B'' are losing and the mixed
time-average game, game, $C= \frac{1}{2}(A+B)$, is winning.
Equation \ref{eq:Parrondo_definition} is satisfied and so we have
Parrondo's ``paradox'' for the two-state games ``A'' and ``B'' as
defined in Equations \ref{eq:A2} and \ref{eq:B2}.
We can simulate the dynamics of this two-state version of Parrondo's
games. Some typical sample paths are shown in Figure \ref{fig:two_state_parrondo}.
The results from the simulations are consistent with the algebraic results.

If we refer back to Figure \ref{fig:two_state_transition} then an
intuitive explanation for this phenomenon is possible. The negative
or ``punishing'' rewards are associated with transitions that do not
change state. The good positive rewards are associated with the
changes of state. If we play a homogeneous sequence of Games ``A' or
``B'' then there are relatively few changes of state and the resulting
weighted sum of all the rewards is negative. If we play the mixed game
then the rewarding changes of state are much more frequent and the
resulting weighted sum of rewards is positive.

\section{Consistency between State-Space and Time averages}
In order for the ``fractal view'' of the process, in state-space, to be
consistent with the time average view of the process we require~:
\begin{equation}
 E \left[ {\bf V_t} \right] = {\bf \Pi_{C} }
\label{eq:consistency}
\end{equation}

The value of $E[ {\bf V_t } ] $ follows from the argument in Section
\ref{sec:Middle-Third}. We can use the mean as defined in
Equation \ref{fractal_mean} to state that
\begin{eqnarray}
   E \left[ {\bf V_t} \right] & = & \left[ \frac{1}{2}+ \frac{1}{2} E \left[ x \right] , 
                         \frac{1}{2} - \frac{1}{2} E \left[ x \right] \right] \\
                   & = & \left[ \frac{1}{2}+ \frac{1}{2} \mu , 
                         \frac{1}{2} - \frac{1}{2} \mu \right] \\
                   & = &  \left[ \frac{1}{2} , \frac{1}{2} \right]
   \label{eq:mean_V}
\end{eqnarray}
The value of ${\bf \Pi_{C} }$ follows from the arguments in Section
\ref{sec:Two-State}. Specifically we require ${\bf \Pi_{C} } = {\bf \Pi_{C} }  \cdot C$ which
gives:
\begin{equation}
{\bf \Pi_{C} } =  \left[ \frac{1}{2} , \frac{1}{2} \right]
\label{eq:Pi_C_value}
\end{equation}
which is consistent with Equation \ref{eq:mean_V}. Which proves this
special case.
To prove the more general case we need to have some notation for 
an entire fractal set, like the one shown in Figure \ref{fig:cantor_fractal}.
We use $ \left\{  F \right\}$ to denote the attractor generated by two operators
$A$ and $B$. We can write~:
\begin{equation}
   E \left[ \left\{ F \right\} \right] =  \gamma  E \left[ \left\{ F \right\} \right] A +
                          \left( 1 - \gamma \right)  E \left[ \left\{ F \right\} \right] B~. 
   \label{eq:proof_of_consistency_1}
\end{equation}
This follows from conservation of measure under the affine transformations
$A$ and $B$. We note that {\bf {\it everything} } in these equations is linear and 
so we can write
\begin{eqnarray}
   E \left[ \left\{ F \right\} \right] & = & E \left[ \left\{ F \right\} \right] 
                          \left( \gamma   A +
                              \left( 1 - \gamma \right) B \right) \\
                       & = & E \left[ \left\{ F \right\} \right] \cdot C 
   \label{eq:proof_of_consistency_2}
\end{eqnarray}
which is the defining property of ${\bf \Pi_{C} }$ which implies that
\begin{equation}
   E \left[ \left\{ F \right\} \right] = {\bf \Pi_{C} }~.
   \label{eq:proof_of_consistency_3}
\end{equation}
The two ways of viewing the situation are consistent
which means that we can use the time averaged game to calculate 
expected values of returns from Parrondo's games. 

\section{Parrondo's original games}

\subsection{Original Definition of Parrondo's Games }

In their original form, Parrondo's games spanned infinite domains, of
all integers or all non-negative integers \cite{Harmer1999a}. 
If our interest is to
examine the asymptotic behaviour of the games as $t \rightarrow \infty$ 
and to study asymptotic rates of return or moments then it is
possible to reduce these games by aggregating states of the Markov
chain modulo three.  We can do this without losing any information
about the rate of return from the games.  After reduction, the Markov
transition operators take the form~:
\begin{equation}
  [A] = \left[
\begin{array}{ccc}
     0       & a_0     & (1-a_0) \\
     (1-a_1) & 0       & a_1     \\
     a_2     & (1-a_2) & 0       \\
\end{array}
  \right]~.
  \label{eq:reduced_matrix}
\end{equation}
where $a_0$, $a_1$ and $a_2$ are the conditional probabilities of
 winning, given the current state modulo three. This form of the 
games has been published by Pearce \cite{Pearce1999b}.

\subsection{Optimised form of Parrondo's Games}

Simulations reveal that {\bf {\it periodic}} inhomogeneous sequences 
of Parrondo's games have the strongest Parrondo effect.
Further investigation by the authors, using Genetic Algorithms,
suggest that the most powerful form of the games is a set of three games
that are played in a strict periodic sequence $\{  G_0 , G_1 , G_2, G_0 , G_1 , G_2, \cdots \}$.
The transition probabilities are as follows~: \\
{\bf Game~$G_0$~:}~$\left[ a_0 , a_1 , a_2  \right] = \left[ \mu       , (1-\mu),  (1-\mu) \right]$ \\
{\bf Game~$G_1$~:}~$\left[ a_0 , a_1 , a_2  \right] = \left[ (1-\mu) , \mu      ,  (1-\mu) \right]$ \\
{\bf Game~$G_2$~:}~$\left[ a_0 , a_1 , a_2  \right] = \left[ (1-\mu) , (1-\mu),  \mu       \right]$ \\
where $\mu$ is a small probability, $0 < \mu < 1$.  We can think
of $\mu$ as being a very small, ideally ``microscopic'', positive
number. The rate of return form any {\bf pure sequence } of these games is
approximately
\begin{equation}
   Y \approx \frac{1}{2} \cdot \mu
   \label{eq:single_rate}
\end{equation}
which is close to zero and yet the return from the 
{\bf cyclic combination} 
of these games is approximately
\begin{equation}
   Y \approx 1 - 3 \cdot \mu
   \label{eq:mixed_rate}
\end{equation}
which is close to a certain win.  We can engineer a situation where we
can deliver an almost certain win every time using games that, on
their own, would deliver almost no benefit at all!  These games
clearly work better as a team than on their own.  Just as team players
may pass the ball in a game of soccer, the games $\{ G_0 , G_1 , G_2 \}$ 
carefully pass the state vector from one trial to the next as this
sequence of Parrondo's games unfolds.

\subsection{An Exquisite Fractal Object}

\begin{figure}[h!]
\hskip.5in   \figeps{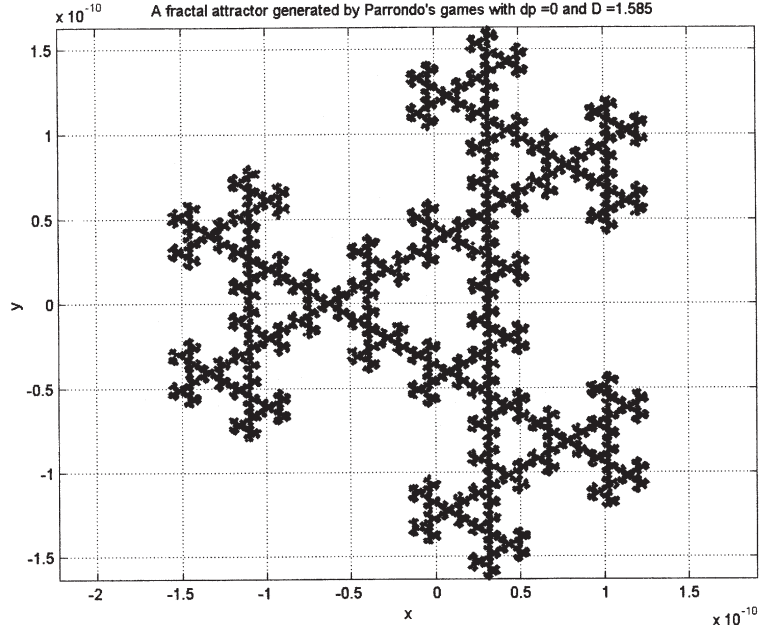}
   \caption{ A 2D projection of a fractal attractor generated by the 
         {\bf ``last gasp''} of Parrondo's games
\label{fig:snowflake} }
\end{figure}

It is possible to de-rate these games by increasing $\mu$. In the
limit as $\mu \rightarrow \frac{1}{2}$ the Parrondo effect vanishes
and the attractor collapses to a single point in state-space. Just
before this limit the attractor takes
the form of the very small and exquisite fractal shown in Figure
\ref{fig:snowflake}.
This fractal is embedded in a two dimensional sub-space of the three
dimensional state-space of the games $\{ G_0 , G_1 , G_2 \}$. The two
dimensional sub-space has been projected onto the page in order to
make it easier to view. The projection preserves dot product, length
and angle measure. The coordinates ``$x$'' and ``$y$'' are linear
combinations of the the components of the original state vector, ${\bf
V_t} = \left[ V_0 , V_1 , V_2 \right] $. The orientation of the image
is such that the original ``$V_2$'' axis is projected onto the new
``$y$'' axis. (The direction of ``up'' is preserved.) The negative
numbers on the axes represent negative offsets rather than negative
probabilities. This is the same concept that is used when we write down
a probability $(1-p)$. If $p$ is a valid probability then so is
$(1-p)$. The number $-p$ is an offset that just happens to be
negative.

The dimension of this fractal is $D \approx \frac{\log (9)} {\log (4)
} \approx 1.585$. We define the amount of Parrondo effect, $\Delta p$,
as the difference in rate of return, $Y$, between the mixed sequence
of games $\{ G_0 , G_1 , G_2 \}$ and the best performance from any
pure sequence of a single game. For this limiting case, $\Delta p
\approx 0$.  There are are some interesting qualitative relationships
between the Hausforff dimension and the amount of Parrondo effect
which deserve further investigation to see if it is possible to 
state a general quantitative law.

\section{Summary}

In this paper we have analysed Parrondo's games in terms of the theory
of Markov chains with rewards. We have illustrated the concepts
constructively, using a very simple two-state version of Parrondo's
games and we have shown how this gives rise to fractal geometry in the
state-space.  We have arrived at a simple method for calculating the
expected value of the asymptotic rate of reward from these games and
we have shown that this can be calculated in terms of an equivalent
time-averaged game.  We have used graphic representations of
trajectories and attractors in state-space to motivate some of the
arguments.

The use of state-space concepts opens up new lines of enquiry.
Simulation and visualisation encourage intuition and help us to grasp
the essential features of a new system. This would be much more
difficult if we were to use a purely formal algebraic approach at the
start. We do not propose visualisation as a {\it replacement} for
rigorous analysis. We see it as a guide to help us to decide which
problems are worthy of more detailed attention and which problems
might later yield to a more formal approach.  We believe that
state-space visualisation will be as useful for the study of the
dynamics of Markov chains as it has already been for the study of
other dynamical systems.

Finally, we conclude that Parrondo's games are not really
``paradoxical'' in the true sense. The anomaly arises because the
reward process is a non-linear function of the Markov transition
operators and our ``common sense'' tells us the reward process
``ought'' to be linear.  When we combine the games by selecting them
at random, we perform a linear convex combination of the operators but
the expected asymptotic value of the rewards from this combined
process is not a linear combination of the rewards from the original
games.

\bibliography{fractal_chapter}

\bibliographystyle{spiebib} 

\end{document}